\documentclass[aps,11pt,nofootinbib,shownopacs,preprintnumbers,prd,superscriptaddress,graphicx]{revtex4}
\usepackage{epsfig,amssymb,amsmath,latexsym,graphics}
\usepackage{subfigure}
\usepackage{caption}
\newcommand\beq{\begin{equation}}
\newcommand\eeq{\end{equation}}
\newcommand\bea{\begin{eqnarray}}
\newcommand\eea{\end{eqnarray}}
\newcommand\bi{\begin{itemize}}
\newcommand\ei{\end{itemize}}
\newcommand\ben{\begin{enumerate}}
\newcommand\een{\end{enumerate}}

\newif\ifboo \boofalse

\def\lsim{\mathrel{\rlap{\lower4pt\hbox{\hskip1pt$\sim$}}
    \raise1pt\hbox{$«$}}}         %less than or approx. symbol
\def\gsim{\mathrel{\rlap{\lower4pt\hbox{\hskip1pt$\sim$}}
    \raise1pt\hbox{$»$}}}         %greater than or approx. symbol
\preprint{CHEP}
\begin{document}
\textheight=23.8cm
\title{\Large{Bounds on fourth generation induced Lepton Flavour Violating double insertions in Supersymmetry}}
\date{\today}
\author{\bf Sumit K. Garg}
\email{sumit@cts.iisc.ernet.in}
\affiliation{Centre for High Energy Physics,
Indian Institute of Science, Bangalore 560 012,
India}%
\author{\bf Sudhir K. Vempati}
\email{vempati@cts.iisc.ernet.in}
\affiliation{Centre for High Energy Physics,
Indian Institute of Science, Bangalore 560 012,
India}

\begin{abstract}
We derive bounds on leptonic double mass insertions
of the type $\delta^{l}_{i4} \delta^{l}_{4j}$ in four generational
MSSM, using the present limits on $l_i \to l_j + \gamma$. 
Two main features distinguish the rates of these processes
in MSSM4 from  MSSM3 : (a) tan$\beta$ is restricted to be
very small $\lesssim 3 $ and (b) the large masses for the
fourth generation leptons.  In spite of small $\tan\beta$, there is
an enhancement in amplitudes with $llrr$($\delta_{i4}^{ll}\delta_{4j}^{rr}$) type insertions   which
pick up the mass of the fourth generation lepton, $m_{\tau'}$. 
We find these bounds to be at least two orders of magnitude
more stringent than those in MSSM3. 
\end{abstract}

\maketitle
\vskip .6 true cm
%------------------------------------------------------------------------
%\section{Introduction}
%\label{intro}
\noindent
\textbf{1}. In the recent times, there has been a renewed interest in the
idea of the fourth generation of Standard Model fermions. While additional
generations have been proposed  quite a while ago \cite{review1,review2}, 
the present  exploration \cite{fourgenstrongdynamics,kribs,Lenzetal,langacker,murayama,
tevatron,tevatronhiggs,hungsher,LHC,
bphysics,cpviolation,dphysics,menkel,buras_lepton,kribs2,sher,ewbaryo,gvw,nandimurdock} 
of the fourth generation is more timely and in tune with 
the on-going searches at LHC\cite{LHC} as well at the Tevatron\cite{Abachi:1995ms,:2008nf}. The presence of
fourth generation can  enhance the production rates of the Higgs at the
Tevatron and thus ruling out a significantly larger mass range\cite{tevatronhiggs} compared
to the three generation SM.

In addition to the direct searches at Colliders\cite{Achard:2001qw,Abachi:1995ms,:2008nf,LHC}, the  fourth generation 
could be probed indirectly in processes where the fourth generation
can contribute through loop effects.    The 
fourth generation contributions to the $S$ and $T$ parameters 
would push them out of the experimentally allowed $(3\sigma)$ range.
A heavy higgs, together with \textit{almost} degenerate masses  for
 the fourth generation has been proposed  by  Kribs \textit{et. al}\cite{kribs} to over 
come this problem (See also \cite{langacker}). 

The presence of the fourth generation would also modify the CKM matrix
thus leading to strong effects in $B$ and $K$ physics. One crucial 
factor is the value of $V_{tb}$. The present Tevatron limits\cite{tevatron}
on $V_{tb}$ would allow it to be as small as $\approx 0.7$ at 3 $\sigma$.  
Such large deviations can lead to significant effects in B-physics \cite{bphysics}. 
Unitarity of the CKM4 together with electroweak precision
measurements  put significant constraints on the allowed forms of the 
four generational CKM matrix.  Similarly, flavour observables in the $K$
and $D$ physics sector would also constraint the mixing and the masses
of the fourth generation quarks\cite{dphysics}.

In the leptonic sector, the effects of fourth generation are quite different
compared to the Standard Model with massive neutrinos.  The fourth
generation neutrino is necessarily greater than 45~$\text{GeV}$ to escape 
the LEP limit on the invisible decay width of $Z$.  As with the CKM,  the PMNS 
matrix is now $4 \times 4 $ whose form determines the couplings of the fourth generation
neutrino. These are strongly constrained from the deviations of the fermi coupling
constant, lepton universality tests as well as rare lepton decays \cite{menkel,buras_lepton}.

Supersymmetric extensions of the four generation can be motivated as a solution
to the little hierarchy problem.  The fourth generation with new additional Yukawa couplings
much larger than the top (especially $t'$ and $b'$ ) can easily enhance the 1-loop
correction to the light higgs mass by a factor of 2 or more. In fact $\text{tan}\beta$ of
$\mathcal{O}(1)$ could now easily be allowed\cite{kribs2}. The relevant Higgs production
cross sections and decay branching fractions have been studied in \cite{sher}. Another
reason why the combination of supersymmetry and four generations is interesting is
that first order phase transition relevant for electroweak baryogenesis is now possible 
without introducing an additional singlet \cite{ewbaryo}.  

Such large Yukawa couplings however elude traditional SUSY-GUT model building
with four generations.  Perturbativity of the Yukawa couplings puts strong UV 
cut-offs on these models, which can utmost be of $\mathcal{O}(100 ~\text{TeV})$ \cite{gvw}. 
Thus perturbative gauge coupling unification is not possible unless additional matter
is added~\cite{nandimurdock}\footnote{Even in the Standard Model,  contrary to the
several statements in the literature, perturbative  unification of the gauge couplings 
at 2-loop is not possible, with the present limits on the fourth generation masses.}.  
In a similar manner, with four generations, it would be hard to realize traditional 
supersymmetric breaking  methods like mSUGRA, minimal AMSB etc with soft terms
defined at the high scale  and renormalisation group evolution determining the 
soft spectrum at the weak  scale.  While minimal gauge mediation is already ruled
out, General Gauge Mediation and variations of it are more suited for the
case of MSSM with four generations\cite{gvw,sumit}.

Similar to the Standard Model with four generations (SM4), one would expect MSSM
with four generations (MSSM4) would also contribute to the flavour processes. However
unlike in the SM4, in MSSM4, flavour violation is determined by the \textit{mis-match} 
between flavour states of SM particles and their super-partners (the super-CKM or 
super-MNS basis).  In fact, it has been known that large flavour violating terms within 
the super-partners are strongly constrained by various flavour violating experiments
\cite{gabbiani}.  One more feature that would make flavour studies within MSSM4
worthwhile is that tan$\beta$ is restricted to be very small $\lesssim 3$.  Thus,  
large $\text{tan}\beta$ enhancement which is typical most MSSM flavour violating
processes, especially the ones which involve dipole operators, is absent within
the case of MSSM4.  Secondly, the large masses of the fourth generation
could lead to enhancement of  amplitudes within the context of some dipole operators. 
Taken together, we think the interplay between two factors make it  worthwhile to 
explore  flavour processes within MSSM4. 

 In the present work, we explore flavour violating constrains in MSSM4. We concentrate on the leptonic sector. 
Typically, the leptonic sector provides an unambiguous constraint  on the 
 flavour violating entries compared to the
hadronic sector where the bounds are dependent on the parameterisation
of the CKM matrix as well as  the uncertainties in the hadronic matrix elements.

Before proceeding further, a couple of comments are in order. The fourth generation neutrino
with a mass $m_{\nu_{\tau^{'}}} > 45$ GeV could be a Majorana particle or a Dirac Particle. While
in the SM, Lepton flavor violation(LFV) processes don't significantly get modified due to this, the construction 
of models in each case could be quite different. In most cases, there could be additional particles at low scale\cite{whepX1}.
In supersymmetric theories, lepton flavour violation is typically proportional  to the scale of supersymmetry breaking. 
 While there could be significant model dependence in construction of the neutrino 
mass matrices, the flavour violation in the supersymmetry breaking soft sector it selves could be a major 
contributing factor.  In the present work, we will assume all the dominant source of LFV in MSSM4 comes 
from the soft sector, which is model independent. In should be noted that in realistic models, in addition 
to the flavour violation from the soft sector, the standard model contribution with four leptonic generations 
and any additional contribution pertaining to the model should be taken in to account.

\begin{figure}[ht]
\includegraphics[width=0.50\textwidth]{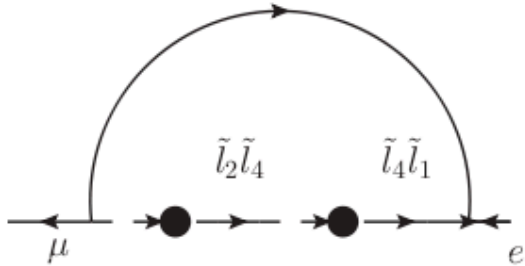}
\caption{{\bf Double Insertions.}
A schematic diagram showing the double insertions of a fourth generation
leading
to flavour violation in 1-2 sector. The photon line is suppressed.
\label{doublemi}}
\end{figure}

\vskip 1cm 
\noindent 
\textbf{2.}  
As is well known, in MSSM,  the dominant source of flavour violation
is from the soft terms.  Thus, in a similar manner to MSSM3,  there are sources
of lepton flavour violation in flavour violating soft terms and are independent of the
neutrino masses. To this extent there could up to twelve new flavour violating 
entries in the soft lagrangian in MSSM4.  These are given as 

\begin{equation}
\mathcal{L}_{\rm soft}  =  \Delta^{i4}_{ll}~ \tilde{l}_i^\star \tilde{l}_4 + \Delta_{rr}^{i4}~ \tilde{r}_i^\star  \tilde{r}_4  + \Delta_{lr}^{i4} ~\tilde{l}_i^\star \tilde{r}_4 + \ldots.. ,
\end{equation}
where $\tilde{l}$ denotes the leptonic doublets (left-handed), $\tilde{r}$ are the leptonic singlets and  
$i = \{1,2,3\}$ for  the standard three generations. While the presence of these terms would definitely 
give rise flavour violating decays for the fourth generation fermions, they would also contribute to 
flavour violating processes in the first three generations. This happens when two fourth generation
flavour violating couplings combine to form a flavour violating entry within the first three generations. 
For illustration purposes, let us consider the case of $\mu \to e + \gamma $.  The diagram 
with fourth generation mass insertions (of the $(ll)$ type )  in shown in Fig.\ref{doublemi}.  
In general this contribution would add to the contribution generated by the flavour violation
already present in the soft potential $\Delta_{ll}^{12}$. 

\noindent
Defining $\delta_{ll}^{ij} ~=~ \Delta_{ll}^{ij} / m_{\tilde{l}}^2$, we can write the
total flavour violating $\delta$ as 
\begin{equation}
\delta_{ll}^{ij} = \delta_{ll}^{ij(3)}  + \delta_{ll}^{ij(4)} 
\end{equation}  
where
\begin{equation}
\delta_{ll}^{ij(4)} = \delta_{ll}^{i4} \delta_{ll}^{4j}
\end{equation}
and $\delta_{ll}^{ij(3)}$ is the which is independent of the presence of the fourth generation. 
These single mass insertions are divided in to four types :$ll$,$rr$,$lr$ and $rl$
depending on the chirality of the corresponding fermion ; it also represents the location of the
flavour violating entry in the slepton mass matrix represented schematically as 
\begin{equation}
\mathcal{M}_{\tilde{f}} = \left( \begin{array}{cc}
m_{\tilde{l} \tilde{l}}^2 & m_{\tilde{l} \tilde{r}}^2 \\
m_{\tilde{r}\tilde{l}}^2 & m_{\tilde{r}\tilde{r}}^2 
\end{array} \right) 
\end{equation}
The possible combinations of double insertions which give the effective single flavour violating
insertions are :
\begin{eqnarray}
l_i l_j = l_i l_4 l_4 l_j  &||& r_i r_j = r_i r_4 r_4 r_j \nonumber \\
l_i r_j  =  l_{i} l_{4} r_{4} r_{j}  \;\; ; \;\;l_i r_4 r_4 r_i  &||&r_i l_j = r_i l_4 l_4 l_j \;\;;\;\; r_i r_4 l_4 l_j  
\label{milist}
\end{eqnarray}
Finally, let us note that $\delta_{ij}^{3}$ can be thought of as being generated by integrating out the 
fourth generation sleptons.  

In the present work, we will derive bounds on the double insertions due to the $\delta_{ij}^{(4)}$.
 In the presence of non-fourth generation flavour violation the bounds
would only become stronger, unless of course one considers fine tuned cancellations between
the two contributions. Thus from now on, we set $\delta_{ij}^{(3)} =0$ and derive the bounds 
on $\delta_{ij}$, where we have suppressed the superscript $(4)$.  We will use the mass insertion
approximations to compute the bounds as done for MSSM3\cite{paride1}. 

Before  we list the amplitudes for each of these mass insertions, let us make a few comments
on the supersymmetric spectrum one considers to evaluate these bounds. For the fourth generation
MSSM, as of now, there is no concrete model of supersymmetry breaking. The classic models of supersymmetry breaking
like minimal supergravity, gauge mediated supersymmetry breaking , Anomaly mediated supersymmetry breaking etc.
which are well established in three generational MSSM cannot be generalized to four generations in their present form\cite{gvw}. 
In all probability\cite{sumit}, the SUSY breaking model could be a  strongly coupled sector  with a low mediation scale in a similar
view to the the general gauge mediation scheme of Seiberg \textit{et. al} \cite{seiberg}. In the following, we will consider a model 
independent approach and evaluate the bounds in generic low energy MSSM. Thus, our approach will be similar to the 
approach taken by Gabbiani  \text{et. al} \cite{gabbiani}.  Accordingly, we will quote our bounds in terms of the slepton 
mass $m_{\tilde{l}(\tilde{r})}$ and  ratios of the parameters  which we are defined 
as  $x_{l(r)}  = M_1^2/m_{{\tilde{l}} ({\tilde{r}})}^2$, $y_{l (r)}  = \mu^2/m_{{\tilde{l}} ({\tilde{r}})}^2$
and $z_{l(r)} = M_2^2/m_{{\tilde{l}}({\tilde{r}})}^2$.  We also fix the ratio $t_{rl} = m_{\tilde{r}}^2 / m_{\tilde{l}}^2$. 
Thus, once the left handed slepton mass and its ratios are given, the right handed slepton mass and its ratios
also get fixed.  A crucial distinction of MSSM4 and MSSM3  is the restriction on tan$\beta$.
 %$x_{l(r)}  = M_1^2/m_{{\tilde{l}} ({\tilde{r}})}^2$, $y_{l (r)}  = \mu^2/m_{{\tilde{l}} ({\tilde{r}})}^2$
%and $x'_{l(r)} = M_2^2/m_{{\tilde{l}}({\tilde{r}})}^2$.
  With the fourth generation masses being very large, tree level
perturbativity restricts tan$\beta$ to be: 
\begin{equation}
\tan\beta \lesssim \left( 2 \pi (v/m_{b'})^2 - 1 \right)^{1/2}
\end{equation}
For a bottom-prime mass, $m_{b'} \approx 300$ GeV, we have 
$\tan\beta ~\approx~ 2$.  This upper bound on tan$\beta$ is
very generic to MSSM4 and is independent of supersymmetric
breaking. It holds as long as one does not change the particle
spectrum. Here we will present our results for few representative 
points in parameter space. In the following we will consider value of tan$\beta$ to be
 2\footnote{Note that such low values are not ruled out by
the light higgs mass constraint in MSSM4.}. The values chosen
for rest of the parameters are given in Table~\ref{paramtable}. 
The parameter space points  S1,S2,S3  are similar to those 
one has in mSUGRA/CMSSM models with universal  scalar and 
gaugino masses at the high scale. (While S1 represents the case
with $m_0 \approx M_{1/2}$, S2 has $M_{1/2} \gg m_0$ and 
S3 represents $m_0 \gg M_{1/2}$.) The point T1, T2 and T3 are motivated
by the general gauge mediation framework of Seiberg \textit{et.al}\cite{seiberg}. The three possible
choices for the ratio $\Lambda_G/\Lambda_S$ (Here $\Lambda_G$, $\Lambda_S$ refers to the gaugino and scalar mass
scale respectively)  fixes $x l$ and $z l$ in this case. $\mu$ is left
to be a free parameter with ratios 0.3, 0.03 and 0.01 for T1, T2 and T3 respectively.

While choosing the points above, we have not taken in to consideration 
the relic density constraints on neutralino dark matter from the WMAP
experiment. The leptonic flavour violating rates would be different
compared to those at the points chosen above. A particularly interesting
case in the three generations  is that of the co-annihilation region 
where the $\tilde{\tau}_1$, has a mass very close to that of the lightest 
neutralino. In MSSM4, it is quite probable that in large regions of the
parameter space,  ${\tilde{\tau}^{'}}_{1}$ is the NLSP. This is especially true
if mSUGRA like boundary conditions could be realized in this model. In such
regions the relation $m_{\tilde{\tau}^{'}} \approx M_{1}$ is roughly satisfied.\\

\begin{table}[h]
\begin{center}
\begin{tabular}{|c|c|c|c|c|}
\hline 
&$x_l$ & $y_l$ & $z_l$ & $t_{rl}$ \\
\hline  
\hline
S1 & 0.1 & 0.3 & 2.5 & 0.7 \\
S2 & 0.3 & 0.1 & 5.5 & 0.4 \\ 
S3 &  0.003 & 0.01 & 0.5 & 0.9 \\
\hline 
\hline
T1 & 0.05 & 0.3 & 0.5 &  0.09\\
T2 & 0.06 & 0.03 & 0.6 &  0.09\\ 
T3 & 0.07  & 0.01 & 0.7 &  0.09\\
\hline
\end{tabular}
\end{center}
\caption{The parameter space points in terms of ratios w.r.t the (left handed ) slepton mass. 
$x_{l}  = M_1^2/m_{\tilde{l}}^2 $, $y_{l}  = \mu^2/m_{{\tilde{l}}}^2$,
$z_{l} = M_2^2/m_{\tilde{l}}^2$, and $t_{rl} = m_{\tilde{r}}^2 / m_{\tilde{l}}^2$.}
\label{paramtable}
\end{table}%

%Three sets corresponds to($\tan\beta=2$ and $m_R= \sqrt{3} m_L$):\\
%\bea
%&(i)& m_L =300, a_{L_1}=0.3, a_{L_2}=0.4, b_L=0.5 \nonumber\\
%&(ii)& m_L =350, a_{L_1}=0.8, a_{L_2}=1.5, b_L=1.8 \nonumber\\
%&(iii)& m_L =400, a_{L_1}=1.2, a_{L_2}=1.7, b_L=2.2 \nonumber\\
%\eea
%

\vskip 1cm 
\noindent 
\textbf{3.}  The contributions from double insertions  of the type $\delta_{ik} \delta_{kj}$ leading to 
flavour violating $i\to j$ processes have already been studied in literature \cite{hisanonomura,paride1}  
for the case of $\mu \to e + \gamma$ where the third generation flavour violation contributes.  
Here we generalize them to the four generation case. 
%The amplitudes for these double insertions contributing to the process $l_i \to l_j + \gamma$, are listed below, where we  have used the following
%notation : $x_{l(r)}  = M_1/m_{{\tilde{l}} ({\tilde{r}})}$, $y_{l (r)}  = \mu/m_{{\tilde{l}} ({\tilde{r}})}$
%and $x'_{l(r)} = M_2/m_{{\tilde{l}}({\tilde{r}})}$. 
We  list below the amplitudes
for the various possible combinations of mass insertions from the fourth generation one by one. 

\begin{figure}
\centering
\mbox{\subfigure{\includegraphics[width=3in]{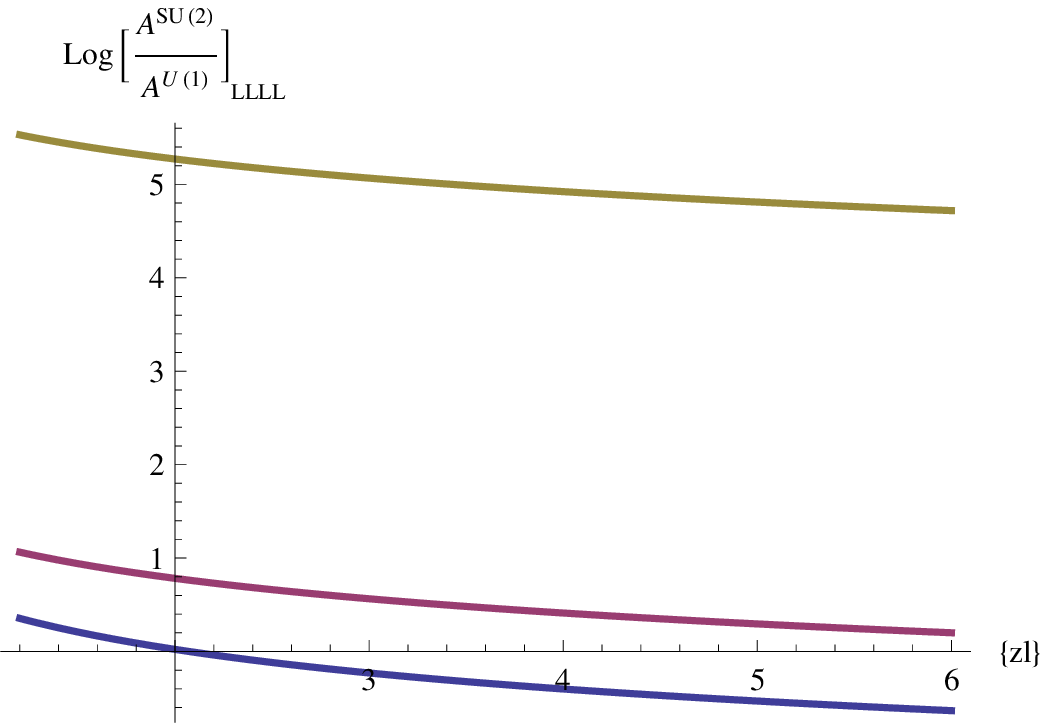}
\quad
\subfigure{\includegraphics[width=3in]{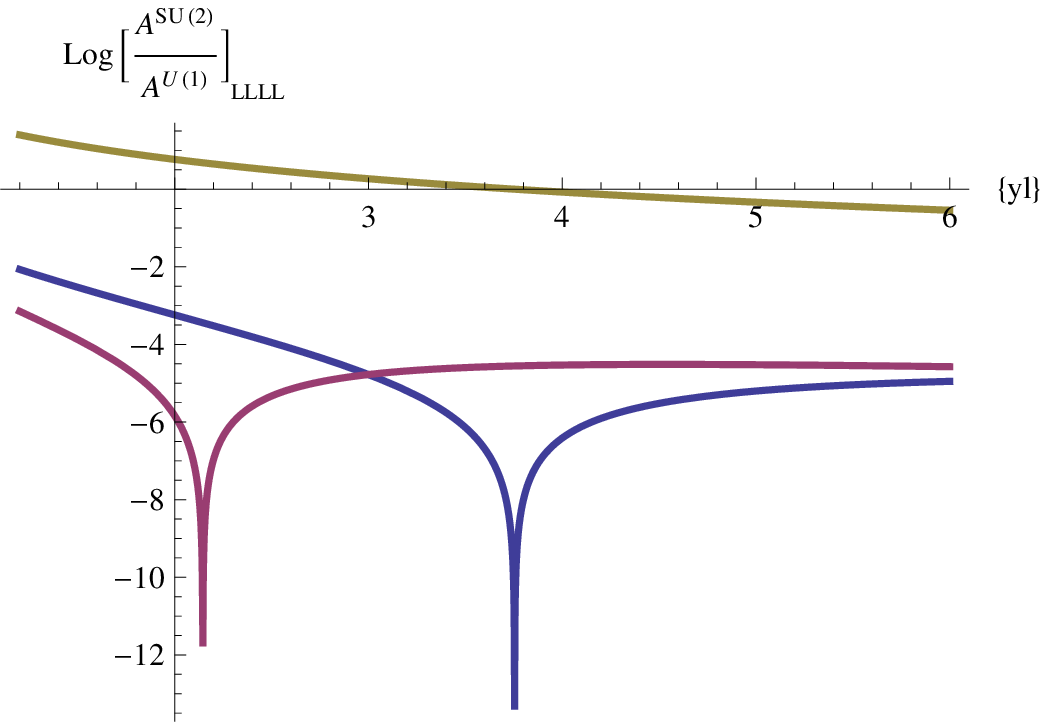} }}}
\caption{Comparison of SU(2) vs U(1) contribution against zl, yl for LLLL case. Rest of the parameter values  correspond to representative points S1, S2 and S3,
 shown by Blue, Pink and 
Brown lines respectively.}
\label{fig12}
\end{figure}

The amplitude associated with $l_i l_4 l_4 l_j$ has contributions
both from chargino as well as neutralino sector. 
These contributions are typically listed as $SU(2)$ and $U(1)$
contributions in the literature \cite{hisanonomura,masina,paride1}: 
\begin{equation}
(A^{ij}_{l2})_{\rm SU(2)} = \tilde{\alpha}_2 \delta^{i4}_{ll} \delta^{4j}_{ll} \left[ { I_{1n} (z_l)  + I_{1c}(z_l)  \over m_{\tilde{l}}^2} + {\mu M_2 \tan\beta \over (M_2^2 -\mu^2)}\left(  { I_{2n}(z_l,y_l) + I_{2c}(z_l,y_l) \over m_{\tilde{l}}^2} \right) \right] 
\end{equation}
\begin{eqnarray}
(A^{ij}_{l2})_{\rm U(1)} &=& \tilde{\alpha}_1 \delta_{ll}^{i4} \delta_{ll}^{4j} \left[ { I_{1n} (x_l)  \over m_{\tilde{l}}^2} + \mu M_1 \tan\beta \left(- { I_{2n} (x_l,y_l) \over m_{\tilde{l}}^2 (M_1^2 - \mu^2) } +
{ 1 \over ( m_{\tilde{r}}^2 - m_{\tilde{l}}^2) } \left(  {2 I_{2n} (x_l) \over m_{\tilde{l}}^2}  \right. \right. \right. \nonumber \\
&& \left. \left. \left. + { 2 f_{2n} (x_l) \over (m_{\tilde{r}}^2 - m_{\tilde{l}}^2) } + { m_{\tilde{l}}^4 \over ( m_{\tilde{r}}^2 - m_{\tilde{l}}^2)^2} \left( {f_{3n} (x_r) \over m_{\tilde{r}}^2} - {f_{3n} (x_l) \over m_{\tilde{l}}^2} \right) \right) \right) \right]  
\end{eqnarray}
Given the larger value of $\tilde{\alpha}_2$ one would expect that the $SU(2)$ contribution to dominate over the $U(1)$ contribution. 
The various loop functions appearing in amplitudes are listed in the appendix.

\begin{figure}
\centering
\mbox{\subfigure{\includegraphics[width=3in]{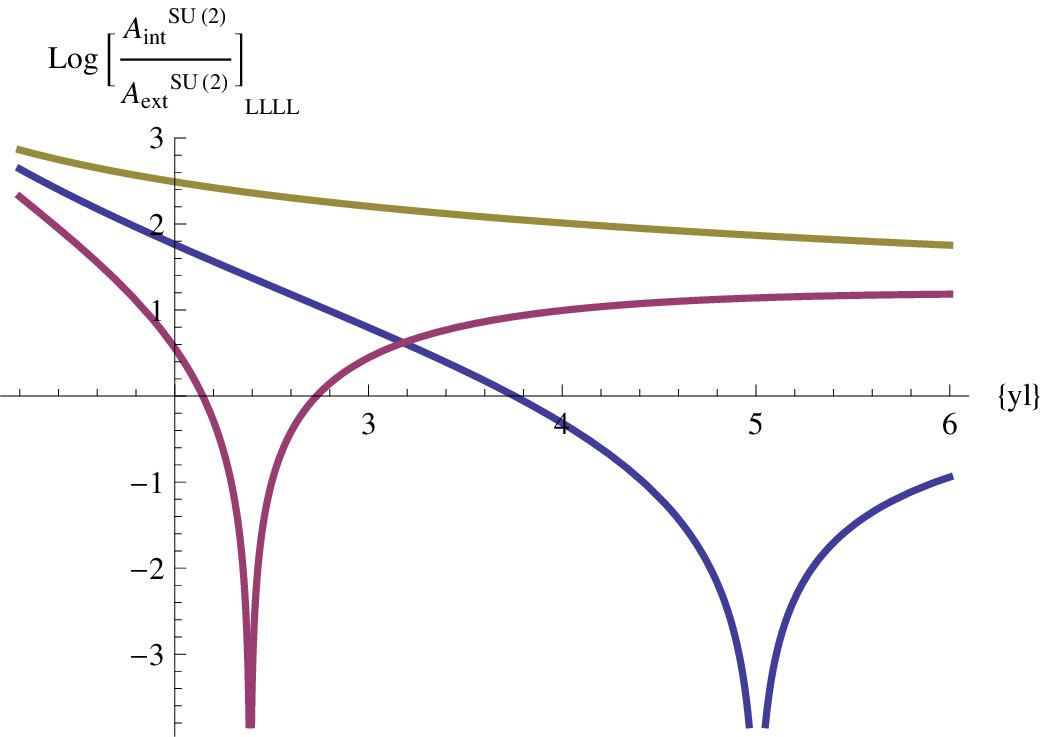}
\quad
\subfigure{\includegraphics[width=3in]{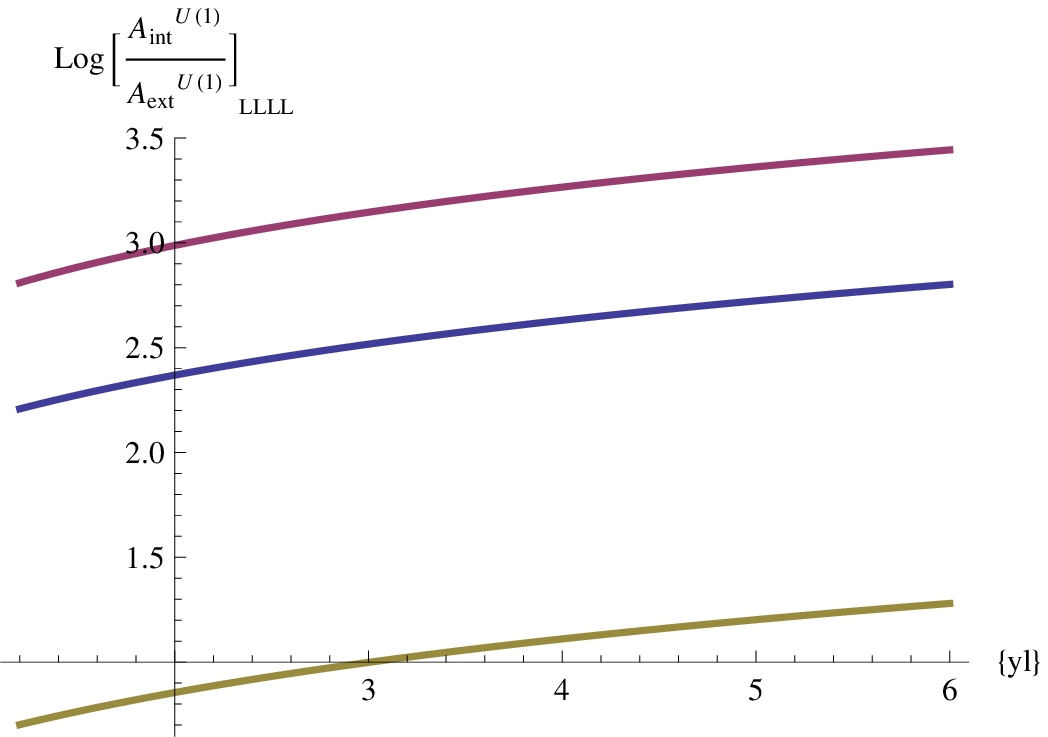} }}}
\caption{Comparison of internal  vs external flip contributions  for SU(2) and U(1) respectively
in LLLL case. Rest of the parameter values  correspond to representative points S1, S2 and S3,
 shown by Blue, Pink and 
Brown lines respectively.}
\label{fig12}
\end{figure}

In Fig. 2 we have shown the comparison of   $SU(2)$ vs. $U(1)$ amplitudes for different parameters sets
taken from  Table I.
It is evident from figures that in some regions of parameter space SU(2) is dominant while in others
U(1) has larger contribution. This dominance is stable under the variation of parameters zl and yl unless there is
cancellation between loop functions. The dips in curves is due to  these cancellations.

Given that $\tan\beta$ is confined to  low values in MSSM4, one would expect that there is no 
large enhancement associated with diagrams with chirality flips either in the vertex
or on the internal line ; both these amplitudes being proportional to $\mu \tan \beta$.
In the allowed regions of $\tan\beta$, the amplitudes of external chirality flip diagrams 
can become comparable in magnitude with those of internal flip ones. This is evident from the
figure (3), where we have have shown the ratios of the internal contribution to the external
contribution for SU(2) and U(1) separately. As can be seen from figure while the internal 
chirality flip diagrams still dominates, there are regions in parameter space where the external
amplitudes become comparable or dominate as can be seen in U(1) contribution for point S3. Of course
there could also be regions where there are cancellations within the internal amplitudes as can be
seen in SU(2) amplitudes for points S2 and S3. Overall we see that for S3, not only U(1) amplitudes 
dominate but also external flip contributions
dominate for small values of $y_l$. 

In the following (Table \ref{LLLL}-\ref{RRRR}) we will present the bounds on the double mass insertions for the spectrum
points $S_i (m_L =200 \text{GeV})$ and $T_i (m_L = 500 \text{GeV})$.
The Branching fraction for $l_i \rightarrow l_j \gamma$ in terms of the amplitudes
is given by \\
\bea
\frac{BR(l_{i}\rightarrow  l_{j}\gamma)}{BR(l_{i}\rightarrow  l_{j}\nu_i\bar{\nu_j})} = 
\frac{48\pi^{3}\alpha}{G_{F}^{2}}(|A_L^{ij}|^2+|A_R^{ij}|^2)
\nonumber
\eea

where $\alpha$ is the fine structure constant and $G_F$ is the Fermi constant.

The present experimental on the limits of the various branching fractions are given as 
\begin{center}
Br($\mu \rightarrow e \gamma$)= $1.2 \times 10^{-11}$ \cite{mega}\\
Br($\tau \rightarrow \mu \gamma$)= $4.4 \times 10^{-8}$\cite{bellebabar}\\
Br($\tau \rightarrow e \gamma$)= $3.3 \times 10^{-8}$\cite{bellebabar}\\
\end{center}

The involved branching ratios of leptonic $\tau$ decays are \cite{pdg}
\begin{center}
Br($\tau \rightarrow \nu_{\tau}\mu\bar{\nu}_{\mu}$)= $(17.36 \pm 0.05)\%$\\
Br($\tau \rightarrow \nu_{\tau}e\bar{\nu}_{e}$)= $(17.84 \pm 0.05)\%$
\end{center}

\begin{table}[htdp]
\begin{center}
\begin{tabular}{|c|c|c|c|}
\hline
MI & $S_1$ & $S_2$ & $S_3$\\
%\hline
  & &  &\\
\hline
$21$ & 0.00114 & 0.00105 & 0.00037 \\
$32$ & 0.16588 & 0.15336 & 0.05469\\
$31$ & 0.14171 & 0.13102 & 0.04672\\
\hline
\hline
MI & $T_1$ & $T_2$ & $T_3$ \\
\hline
$21$ & 0.00237 & 0.00215 & 0.00242 \\
$32$ & 0.34525 & 0.31317 & 0.35238\\
$31$ & 0.29495 & 0.26754 & 0.30103\\
\hline
\end{tabular}
\end{center}
\caption{Bounds on (($\delta_{LL})_{i4} (\delta_{LL} )_{4j}$) }
\label{LLLL}
\end{table}%

\begin{figure}
\centering
\mbox{\subfigure{\includegraphics[width=3.2in]{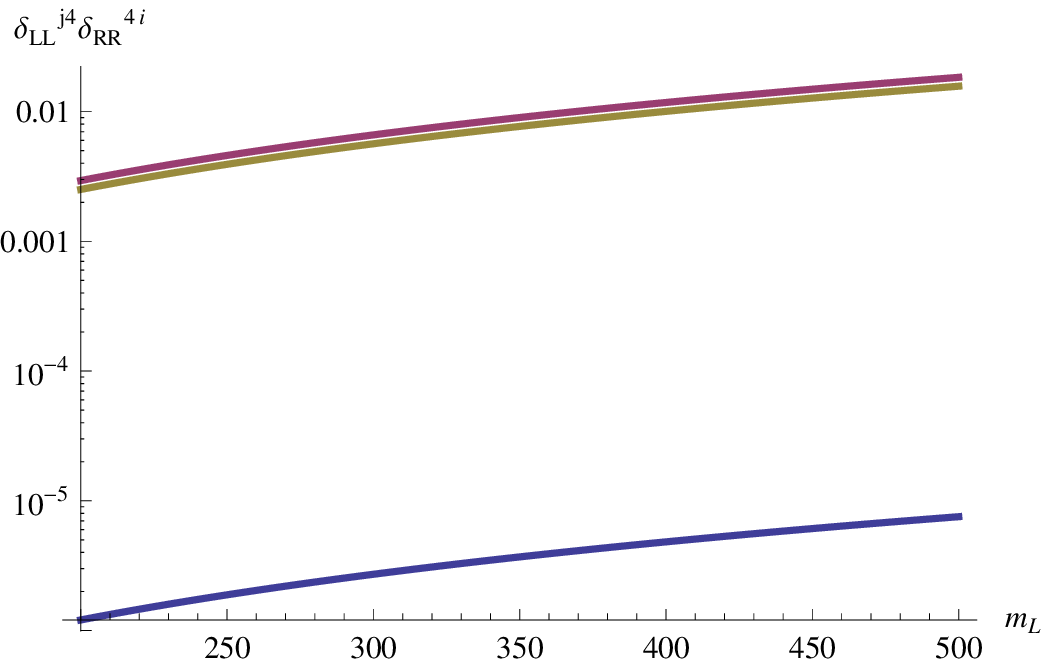}
\quad
\subfigure{\includegraphics[width=3.2in]{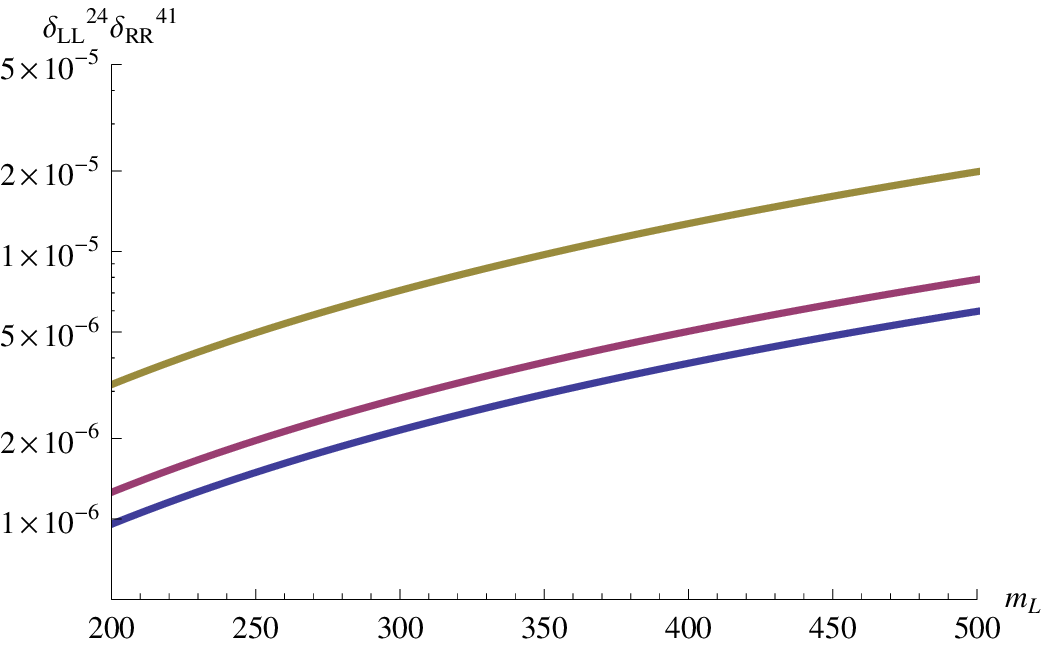} }}}
\vspace{-2cm}
\caption{Variation of $\delta_{LL}\delta_{RR}$ bound  w.r.t slepton mass i.e. $m_L$.
Ist figure corresponds to bound in 21(blue), 32(pink) and 31(brown) sector while the 2nd 
figure is with different trl values(blue, pink and brown corresponds to $t_{rl}$ values of 0.01, 0.1 and 0.5
respectively) for 21 case. Merging of pink and brown line is due to nearly equal bound in 23 and
31 sector.}\label{fig12}
\end{figure}

\vskip 1cm 
\noindent 
\textbf{4.} The chirality flip associated with the fourth generation lepton mass however, makes its appearance in amplitudes with 
double mass insertions of the type $l_i l_4 r_4 r_j$ where there would be an chirality flipping 
$m^2_{\tilde{l}_4 \tilde{r}_4} = m_{\tau'} \mu \tan \beta$ mass insertion.  Here the amplitude gets enhanced by a 
$m_{\tau'}/m_i$ factor associated with the mass of the decaying lepton.  This factor which could be quite large
could significantly strengthen the bounds by an order of magnitude or more, depending on the mass of the 
fourth generation lepton ($\tau'$) chosen.  The amplitude for this mass insertion is given by:
 
\begin{eqnarray}
A^{ij}_{l3} &= &-2 \tilde{\alpha}_1  {m_{\tau'} \over m_i} \mu M_1 \tan\beta \delta^{i4}_{ll} \delta^{4j}_{rr} 
{ m_{\tilde{l}}^2 m_{\tilde{r}}^2 \over (m_{\tilde{l}}^2 - m_{\tilde{r}}^2)^2 } \left( {f_{2n} (x_l)  \over m_{\tilde{l}}^4}  
+ { f_{2n}(x_r) \over m_{\tilde{r}}^4} \right. \nonumber \\  &&\left. + {1 \over (m_{\tilde{r}}^2 - m_{\tilde{l}}^2) }
 \left[ {f_{3n} (x_r)  \over m_{\tilde{r}}^2 } - { f_{3n} (x_l) \over m_{\tilde{l}}^2 } \right] \right)   
\end{eqnarray}

For present work we have chosen $m_{\tau'} = 100 ~\text{GeV}$ consistent with present limits from direct searches \cite{pdg}. 
In Fig. 4 we have shown the variation of LLRR bound w.r.t slepton mass, $m_L$. The bound scales  
inversely with increasing value of (square of) $m_L$. The 
bound also becomes weaker as move to higher values of $m_R$ as evident 
from second part of Fig. 4 with different
$t_{rl}$ values. As mentioned previously the bounds on LLRR are sensitive to $m_{\tau^{'}}$ and thus they
have stronger constraints compared to other double insertions. In Tables III, IV we present
a comparison of bounds in MSSM4 and MSSM3. It is clear MSSM4 bounds are much stronger, by atleast
couple of orders of magnitude.\\ 

\begin{minipage}[b]{.40\textwidth}
  \centering
  \begin{tabular}{|c|c|c|c|}
\hline
MI & $S_1$ & $S_2$ & $S_3$\\
%\hline
  & & & \\
\hline
$21$ & $3.97 \times 10^{-6}$& $6.46 \times 10^{-6}$ & $6.75 \times 10^{-5}$\\
$32$ & 0.00969 & 0.01576 & 0.16447\\
$31$ & 0.00828 & 0.01346 & 0.14051\\
\hline
\hline
MI & $T_1$ & $T_2$ & $T_3$\\
\hline
$21$ & $7.54 \times 10^{-6}$& $2.49 \times 10^{-5}$ & $4.51 \times 10^{-5}$\\
$32$ & 0.01837 & 0.06076 & 0.11004\\
$31$ & 0.01570 & 0.05191 & 0.09401\\
\hline
\end{tabular}
  \captionof{table}{Bounds on (($\delta_{LL})_{i4} (\delta_{RR} )_{4j}$) for $m_{\tau'} = 100 ~\text{GeV}$ from MSSM4. It varies linearly with inverse of $m_{\tau'}$.}
  \label{LLRR4}
\end{minipage}\qquad
\begin{minipage}[b]{.40\textwidth}
  \centering
  \begin{tabular}{|c|c|c|c|}
\hline
MI & $S_1$ & $S_2$ & $S_3$\\
%\hline
  & & & \\
\hline
$21$ & $2.23 \times 10^{-4}$& $3.64 \times 10^{-4}$ & $0.00379$\\
$32$ & 0.54556 & 0.88706 & -\\
$31$ & 0.46607 & 0.75781 & -\\
\hline
\hline
MI & $T_1$ & $T_2$ & $T_3$\\
\hline
$21$ & $4.24 \times 10^{-4}$& $1.40 \times 10^{-3}$ & $2.54 \times 10^{-3}$\\
$32$ & 1.03418 & - & -\\
$31$ & 0.88349 & - & -\\
\hline
\end{tabular}
  \captionof{table}{Bounds on (($\delta_{LL})_{i3} (\delta_{RR} )_{3j}$) from MSSM3. Hyphen(-) sign indicates the unphysical bound larger than unity. }
  \label{LLRR3}
\end{minipage}

The amplitudes associated with $rlll$ double insertions is given as follows. The corresponding bounds
are presented in Table V.
\begin{equation}
A_{l1}^{ij} = \tilde{\alpha}_1 {M_1 \over m_i}  \delta^{i4}_{rl} \delta^{4j}_{ll}  { m_{\tilde{l}}^2 
m_{\tilde{r}}^2  \over  (m_{\tilde{l}}^2 - m_{\tilde{r}}^2) } \left[ { 2 f_{2n}(x_l)  \over m_{\tilde{l}}^4}+ 
{ 1  \over (m_{\tilde{l}}^2 - m_{\tilde{r}}^2)} \left( { f_{3n} (x_l) \over m_{\tilde{l}}^2 } -  { f_{3n}(x_r) \over m_{\tilde{r}}^2 } \right) \right]
\end{equation}   
The amplitude associated with $(lr)(rr)$ is given by the above expression with ($l \leftrightarrow r $)(for corresponding bounds see Tabel \ref{LRRR}). 

\begin{table}[htdp]
\begin{center}
\begin{tabular}{|c|c|c|c|}
\hline
MI & $S_1$ & $S_2$ & $S_3$\\
%\hline
  & &  &\\
\hline
$21$ & $2.57 \times 10^{-6}$ & $3.29 \times 10^{-6}$ & $7.36 \times 10^{-6}$\\
$32$ & $0.00626$ & $0.00801$ & $0.01794$\\
$31$ & $0.00535$ & $0.00685$ & $0.01532$\\
\hline
\hline
MI & $T_1$ & $T_2$ & $T_3$\\
\hline
$21$ & $1.14 \times 10^{-5}$ & $1.15 \times 10^{-5}$ & $1.16 \times 10^{-5}$\\
$32$ & $0.02787$ & $0.02804$ & $0.02834$\\
$31$ & $0.02381$ & $0.02390$ & $0.02421$\\
\hline
\end{tabular}
\end{center}
\caption{Bounds on (($\delta_{RL})_{i4} (\delta_{LL} )_{4j}$) }
\label{RLLL}
\end{table}%

\begin{table}[htdp]
\begin{center}
\begin{tabular}{|c|c|c|c|}
\hline
MI & $S_1$ & $S_2$ & $S_3$\\
%\hline
  & &  &\\
\hline
$21$ & $1.89 \times 10^{-6}$ & $1.67 \times 10^{-6}$ & $6.63 \times 10^{-6}$\\
$32$ & $0.00462$  & $0.00408$ & $0.01617$ \\
$31$ & $0.00395$  & $0.00348$ & $0.01381$ \\
\hline
\hline
MI & $T_1$ & $T_2$ & $T_3$\\
\hline
$21$ & $1.56 \times 10^{-6}$ & $1.62 \times 10^{-6}$ & $1.69 \times 10^{-6}$\\
$32$ & $0.00381$  & $0.00397$ & $0.00413$ \\
$31$ & $0.00326$  & $0.00339$ & $0.00353$ \\
\hline
\end{tabular}
\end{center}
\caption{Bounds on (($\delta_{LR})_{i4} (\delta_{RR} )_{4j}$) }
\label{LRRR}
\end{table}%

Finally, the amplitude associated with $rrrr$ double mass insertions is given by 

\begin{eqnarray}
A^{ij}_{r2} &=& \tilde{\alpha}_1 \delta^{i4}_{rr} \delta^{4j}_{rr} \left[ 4 { I_{1n} (x_r) \over m_{\tilde{r}}^2} + \mu \tan\beta M_1 \left( 2 { I_{2n} (x_r,y_r ) \over 
m_{\tilde{r}}^2 (M_1^2 - \mu^2) } \right. \right. \nonumber \\
&& \left. \left. + {1 \over (m_{\tilde{r}}^2 - m_{\tilde{l}}^2)} \left\{ - 2 {  I_{2n} (x_r) \over m_{\tilde{r}}^2} + 2 { f_{2n}(x_r)  \over (m_{\tilde{r}}^2 - m_{\tilde{l}}^2)} + 
{ m_{\tilde{r}}^4 \over (m_{\tilde{r}}^2 - m_{\tilde{l}}^2)^2} \left( {f_{3n} (x_r) \over m_{\tilde{r}}^2} - {f_{3n}(x_l) \over m_{\tilde{l}}^2} \right) \right\} \right) \right]   
\end{eqnarray}

\begin{table}[htdp]
\begin{center}
\begin{tabular}{|c|c|c|c|}
\hline
MI & $S_1$ & $S_2$ & $S_3$\\
%\hline
  & & & \\
\hline
$21$ & 0.00113& 0.00081 & 0.00139\\
$32$ & 0.16521 & 0.1189 &  0.20335\\
$31$ & 0.14113 & 0.10157 &  0.17372\\
\hline
\hline
MI & $T_1$ & $T_2$ & $T_3$\\
\hline
$21$ & 0.00191& 0.00104 & 0.00106\\
$32$ & 0.27888 & 0.15229 & 0.15447\\
$31$ & 0.23825 & 0.13010 & 0.13196\\
\hline
\end{tabular}
\end{center}
\caption{Bounds on (($\delta_{RR})_{i4} (\delta_{RR} )_{4j}$) }
\label{RRRR}
\end{table}%

The corresponding bounds on double insertions are given in Table VII. As one can see
like MSSM3 the constraints 
on these parameters are very weak in this case.

\vskip 0.5cm 
\noindent
\textbf{5}. 
%In addition to the standard neutralino and chargino contributions considered
%above, the Higgs bosons can also contribute to the flavour violating processes and these
%contributions could become important in some regions of the parameter space. In the following,
%we comment on this aspect. 
%\section{Discussion}
Double insertions are an effective way of constraining four generation
flavour violating entries in supersymmetric theories. The importance
of these insertions has already been stressed in the works of Hisano
et.al\cite{hisanonomura} and Paradisi\cite{paride1}. In the present work, we 
have used this approach to constraint fourth generation flavour
violating entries from the existing lepton flavour violating decays. 
While most chiral combinations of these entries like LLLL or RRRR
etc have bounds similar to that of the single insertions, LLRR 
insertions are special as they pick up the mass of the fourth 
generation lepton leading to enhanced amplitudes. The resultant
bounds are stringent by at least an order of magnitude and could
reach up to three orders of magnitude stronger constraints compared
to the existing ones. Of course, please note that these are just
conservative bounds in the limit the single insertions are negligible;
in their presence the bounds are further stringent. 

In the present work, we have considered constraints only
from the lepton flavour violating decays considering dipole 
transitions from gauge interactions. In addition to these processes,
the double insertions could play a role in EDMs also\cite{hisanoparide}.
The large mass of the fourth generation particle can lead to enhanced
contributions to the EDMs. Similarly, Higgs mediated diagrams
\cite{babu-kolda,anna-andrea,paride-higgs} could have transitions 
with double insertions. The LLRR insertion as in the present case 
could have enhanced contribution due to the large fourth generation 
mass insertion compared to its third generation counterpart, however they
may be suppressed due to the low $\tan\beta$ requirement of MSSM4. The 
interplay between these two effects need to be explored.

%-------------------------------------------------------------------------
\acknowledgements{}
 SKV is supported by DST Grant, Govt. of India , ``Complementarity between 
 direct and Indirect searches of Supersymmetry" and by a  DST Ramanujan Fellowship(SR/S2/RJN-25/2008).  
 The work of S.K.G is also partially supported by this DST Ramanujan Fellowship. 

\newpage
{\bf{Appendix}} : {\bf{Loop Functions}}
\vskip .5 true cm

In this appendix we will give the explicit form of loop functions appearing in
amplitudes:

\bea
 f_{2n}(x) &=& \frac{-5x^2+4x+1+2x(x+2)\ln x}{4(1-x)^4}\nonumber\\
f_{3n}(x) &=& \frac{1+2x\ln x-x^2}{2( 1-x)^3}\nonumber\\
  I_{1n}(x) &=& \frac{3x^4+44x^3-36x^2-12x+1-12x^2(2x+3)\ln x}{24(1-x)^6}\nonumber\\
 I_{2n}(x)&=&\frac{x^3+9x^2-9x-1-6x(x+1)\ln x}{4(1-x)^5}\nonumber\\
 I_{1c}(x)&=&\frac{10x^3+9x^2-18x-1-3x(3+6x+x^2)\ln x}{6(1-x)^6}\nonumber\\
 I_{2c}(x)&=&\frac{3x^2-3-(x^2+4x+1)\ln x}{(1-x)^5}  \nonumber\\
I_{2(c,n)}(x,y)&=& I_{2(c,n)}(x) - I_{2(c,n)}(y).\nonumber\\
\eea

%%%%%%%%%%%%%%%%%%%%%%%
%\bibliographystyle{apsrev}

%%%%%%%%%%%%%%%%%%%%%%%%


\begin{thebibliography}{99}

%%%%% Introduction %%%%%%%%%

\bibitem{review1}
For original references and a review,  please see, \\
%\cite{Frampton:1999xi}
%\bibitem{Frampton:1999xi}
  P.H.~Frampton, P.~Q.~Hung and M.~Sher,
  %``Quarks and leptons beyond the third generation,''
  Phys.\ Rept.\  {\bf 330}, 263 (2000)
  [arXiv:hep-ph/9903387];
  %%CITATION = PRPLC,330,263;%%
  %%
  \bibitem{review2}
 For  a recent review, please see
  %\cite{Holdom:2009rf}
%\bibitem{Holdom:2009rf}
  B.~Holdom, W.~S.~Hou, T.~Hurth, M.~L.~Mangano, S.~Sultansoy and G.~Unel,
  %``Four Statements about the Fourth Generation,''
  PMC Phys.\  A {\bf 3}, 4 (2009)
  [arXiv:0904.4698 [hep-ph]].
  %%CITATION = PMCPA,A3,4;%%
  
  
\bibitem{fourgenstrongdynamics}
B. Holdom, Phys. Rev. Lett. {\bf 57}, 2496 (1986)
[Erratum-ibid. {\bf 58}, 177 (1987);
W.A. Bardeen, C.T. Hill and M. Lindner, Phys. Rev. {\bf D41}, 1647 (1990);
C. Hill, M. Luty and E.A. Paschos, Phys. Rev. {\bf D43}, 3011 (1991);
P.Q. Hung and G. Isidori Phys. Lett. {\bf B402}, 122 (1997);
%%
%\cite{He:2001fz}
%\bibitem{He:2001fz}
  H.~J.~He, C.~T.~Hill and T.~M.~P.~Tait,
  %``Top quark seesaw, vacuum structure and electroweak precision
  %constraints,''
  Phys.\ Rev.\  D {\bf 65}, 055006 (2002)
  [arXiv:hep-ph/0108041];
  %%CITATION = PHRVA,D65,055006;%%
%%
%\cite{Hashimoto:2009ty}
%%\bibitem{Hashimoto:2009ty}
  M.~Hashimoto and V.~A.~Miransky,
  %``Dynamical electroweak symmetry breaking with superheavy quarks and 2+1
  %composite Higgs model,''
  Phys.\ Rev.\  D {\bf 81}, 055014 (2010)
  [arXiv:0912.4453 [hep-ph]].
  %%CITATION = PHRVA,D81,055014;%%
%%
%\cite{Frandsen:2009fs}
%\bibitem{Frandsen:2009fs}
  M.~T.~Frandsen, I.~Masina and F.~Sannino,
  %``Fourth Lepton Family is Natural in Technicolor,''
  Phys.\ Rev.\  D {\bf 81}, 035010 (2010)
  [arXiv:0905.1331 [hep-ph]].
  %%CITATION = PHRVA,D81,035010;%%
  %
%\cite{Hung:2009ia}
%\bibitem{Hung:2009ia}
  P.~Q.~Hung and C.~Xiong,
  %``Implication of a Quasi Fixed Point with a Heavy Fourth Generation: The
  %emergence of a TeV-scale physical cutoff,''
  Phys.\ Lett.\  B {\bf 694}, 430 (2011)
  [arXiv:0911.3892 [hep-ph]];
  %%CITATION = PHLTA,B694,430;%%
%\cite{Hung:2009hy}
%\bibitem{Hung:2009hy}
  P.~Q.~Hung and C.~Xiong,
  %``Renormalization Group Fixed Point with a Fourth Generation: Higgs-induced
  %Bound States and Condensates,''
  Nucl.\ Phys.\  B {\bf 847}, 160 (2011)
  [arXiv:0911.3890 [hep-ph]].
  %%CITATION = NUPHA,B847,160;%%
%%
%\cite{Hung:2010xh}
%\bibitem{Hung:2010xh}
  P.~Q.~Hung and C.~Xiong,
  %``Dynamical Electroweak Symmetry Breaking with a Heavy Fourth Generation,''
  arXiv:1012.4479 [hep-ph];
  %%CITATION = ARXIV:1012.4479;%%
%\cite{Delepine:2010vw}
%\bibitem{Delepine:2010vw}
  D.~Delepine, M.~Napsuciale and C.~A.~Vaquera-Araujo,
  %``Dynamical symmetry breaking with a fourth generation,''
  arXiv:1003.3267 [hep-ph].
  %%CITATION = ARXIV:1003.3267;%%
%%
%\cite{Fukano:2011fp}
%\bibitem{Fukano:2011fp}
  H.~S.~Fukano and K.~Tuominen,
  %``Topcolor-like dynamics and new matter generations,''
  arXiv:1102.1254 [hep-ph];
  %%CITATION = ARXIV:1102.1254;%%
%\cite{Ho:2011qi}
%\bibitem{Ho:2011qi}
  C.~M.~Ho, P.~Q.~Hung and T.~W.~Kephart,
  %``Conformal Completion of the Standard Model with a Fourth Generation,''
  arXiv:1102.3997 [hep-ph].
  %%CITATION = ARXIV:1102.3997;%%





\bibitem{LHC}
%\cite{Cacciari:2008zb}
%\bibitem{Cacciari:2008zb}
  M.~Cacciari, S.~Frixione, M.~L.~Mangano, P.~Nason and G.~Ridolfi,
  %``Updated predictions for the total production cross sections of top and of
  %heavier quark pairs at the Tevatron and at the LHC,''
  JHEP {\bf 0809}, 127 (2008)
  [arXiv:0804.2800 [hep-ph]]; 
  %%CITATION = JHEPA,0809,127;%%
  %\cite{Dobrescu:2009vz}
%%\bibitem{Dobrescu:2009vz}
  B.~A.~Dobrescu, K.~Kong and R.~Mahbubani,
  %``Prospects for top-prime quark discovery at the Tevatron,''
  JHEP {\bf 0906}, 001 (2009)
  [arXiv:0902.0792 [hep-ph]];
  %%CITATION = JHEPA,0906,001;%% 
%%
%\cite{Rajaraman:2010ua}
%\bibitem{Rajaraman:2010ua}
  A.~Rajaraman and D.~Whiteson,
  %``Tevatron Discovery Potential for Fourth Generation Neutrinos: Dirac,
  %Majorana and Everything in Between,''
  Phys.\ Rev.\  D {\bf 82}, 051702 (2010)
  [arXiv:1005.4407 [hep-ph]].
  %%CITATION = PHRVA,D82,051702;%%
%%
%\cite{Das:2010fh}
%\bibitem{Das:2010fh}
  D.~Das, D.~London, R.~Sinha and A.~Soffer,
  %``Measuring the Magnitude of the Fourth-Generation CKM4 Matrix Element
  %V_{t'b'},''
  Phys.\ Rev.\  D {\bf 82}, 093019 (2010)
  [arXiv:1008.4925 [hep-ph]].
%%  
  %%CITATION = PHRVA,D82,093019;%%
%%
%\cite{Choudhury:2010ya}
%\bibitem{Choudhury:2010ya}
  D.~Choudhury and D.~K.~Ghosh,
  %``A fourth generation, anomalous like-sign dimuon charge asymmetry and the
  %LHC,''
  JHEP {\bf 1102}, 033 (2011)
  [arXiv:1006.2171 [hep-ph]];
  %%CITATION = JHEPA,1102,033;%%
  %%
%\cite{Carpenter:2010bs}
%\bibitem{Carpenter:2010bs}
  L.~M.~Carpenter, A.~Rajaraman and D.~Whiteson,
  %``Searches for Fourth Generation Charged Leptons,''
  arXiv:1010.1011 [hep-ph].
  %%CITATION = ARXIV:1010.1011;%%
%%
%\cite{BarShalom:2010bh}
%\bibitem{BarShalom:2010bh}
  S.~Bar-Shalom, G.~Eilam and A.~Soni,
  %``Collider signals of a composite Higgs in the Standard Model with four
  %generations,''
  Phys.\ Lett.\  B {\bf 688}, 195 (2010)
  [arXiv:1001.0569 [hep-ph]].
  %%CITATION = PHLTA,B688,195;%%





\bibitem{tevatronhiggs}
%\cite{Aaltonen:2010sv}
%\bibitem{Aaltonen:2010sv}
  T.~Aaltonen {\it et al.}  [CDF and D0 Collaboration],
  %``Combined Tevatron upper limit on gg->H->W+W- and constraints on the Higgs
  %boson mass in fourth-generation fermion models,''
  arXiv:1005.3216 [hep-ex].
  %%CITATION = ARXIV:1005.3216;%%
The modified Higgs branching fractions can be found in : 
%\cite{Belotsky:2002ym}
%\bibitem{Belotsky:2002ym}
  K.~Belotsky, D.~Fargion, M.~Khlopov, R.~Konoplich and K.~Shibaev,
  %``Invisible Higgs boson decay into massive neutrinos of 4th generation,''
  Phys.\ Rev.\  D {\bf 68}, 054027 (2003)
  [arXiv:hep-ph/0210153].
  %%CITATION = PHRVA,D68,054027;%%






\bibitem{kribs}
G.~D.~Kribs, T.~Plehn, M.~Spannowsky and T.~M.~P.~Tait,
  %``Four generations and Higgs physics,''
  Phys.\ Rev.\  D {\bf 76}, 075016 (2007)
  [arXiv:0706.3718 [hep-ph]].
  %%CITATION = PHRVA,D76,075016;%%



%\cite{Bobrowski:2009ng}
\bibitem{Lenzetal}
  M.~Bobrowski, A.~Lenz, J.~Riedl and J.~Rohrwild,
  %``How Much Space Is Left For A New Family Of Fermions?,''
  Phys.\ Rev.\  D {\bf 79}, 113006 (2009)
  [arXiv:0902.4883 [hep-ph]].
  %%CITATION = PHRVA,D79,113006;%%




\bibitem{langacker}
%\cite{Erler:2010sk}
%\bibitem{Erler:2010sk}
  J.~Erler and P.~Langacker,
  %``Precision Constraints on Extra Fermion Generations,''
  Phys.\ Rev.\ Lett.\  {\bf 105}, 031801 (2010)
  [arXiv:1003.3211 [hep-ph]].
  %%CITATION = PRLTA,105,031801;%%

\bibitem{murayama}
 H.~Murayama, V.~Rentala, J.~Shu and T.~T.~Yanagida,
  %``Saving fourth generation and baryon number by living long,''
  arXiv:1012.0338 [hep-ph].
  %%CITATION = ARXIV:1012.0338;%%

\bibitem{tevatron}
 %\cite{Aaltonen:2009nr}
%\bibitem{Aaltonen:2009nr}
  T.~Aaltonen {\it et al.}  [CDF Collaboration],
  %``Search for New Bottomlike Quark Pair Decays $Q\Qbar \to
  %(\t\Wmp)(\tbar\Wpm)$ in Same-Charge Dilepton Events,''
  Phys.\ Rev.\ Lett.\  {\bf 104}, 091801 (2010)
  [arXiv:0912.1057 [hep-ex]].
  %%CITATION = PRLTA,104,091801;%%


\bibitem{hungsher}
%\cite{Hung:2007ak}
%\bibitem{Hung:2007ak}
  P.~Q.~Hung and M.~Sher,
  %``Experimental constraints on fourth generation quark masses,''
  Phys.\ Rev.\  D {\bf 77}, 037302 (2008)
  [arXiv:0711.4353 [hep-ph]].
  %%CITATION = PHRVA,D77,037302;%%
%\cite{Flacco:2010rg}
%%%

%\bibitem{Flacco:2010rg}
  C.~J.~Flacco, D.~Whiteson, T.~M.~P.~Tait and S.~Bar-Shalom,
  %``Direct Mass Limits for Chiral Fourth-Generation Quarks in All Mixing
  %Scenarios,''
  Phys.\ Rev.\ Lett.\  {\bf 105}, 111801 (2010)
  [arXiv:1005.1077 [hep-ph]].
  %%CITATION = PRLTA,105,111801;%%



\bibitem{bphysics}

%\cite{Alwall:2006bx}
%\bibitem{Alwall:2006bx}
  J.~Alwall {\it et al.},
  %``Is V(tb) ~= 1?,''
  Eur.\ Phys.\ J.\  C {\bf 49} (2007) 791
  [arXiv:hep-ph/0607115].
  %%CITATION = EPHJA,C49,791;%%

%\cite{Soni:2008bc}
%\bibitem{Soni:2008bc}
  A.~Soni, A.~K.~Alok, A.~Giri, R.~Mohanta and S.~Nandi,
  %``The Fourth family: A Natural explanation for the observed pattern of
  %anomalies in $B^-$ CP asymmetries,''
  Phys.\ Lett.\  B {\bf 683}, 302 (2010)
  [arXiv:0807.1971 [hep-ph]].
  %%CITATION = PHLTA,B683,302;%%
%%
%\cite{Soni:2010xh}
%\bibitem{Soni:2010xh}
  A.~Soni, A.~K.~Alok, A.~Giri, R.~Mohanta and S.~Nandi,
  %``SM with four generations: Selected implications for rare B and K decays,''
  Phys.\ Rev.\  D {\bf 82}, 033009 (2010)
  [arXiv:1002.0595 [hep-ph]];
  %%CITATION = PHRVA,D82,033009;%%
%%
%\cite{Buras:2010pi}
%\bibitem{Buras:2010pi}
  A.~J.~Buras, B.~Duling, T.~Feldmann, T.~Heidsieck, C.~Promberger and S.~Recksiegel,
  %``Patterns of Flavour Violation in the Presence of a Fourth Generation of
  %Quarks and Leptons,''
  JHEP {\bf 1009}, 106 (2010)
  [arXiv:1002.2126 [hep-ph]];
  %%CITATION = JHEPA,1009,106;%%
%%

%%
O.~Eberhardt, A.~Lenz and J.~Rohrwild,
  %``Less space for a new family of fermions,''
  Phys.\ Rev.\  D {\bf 82}, 095006 (2010)
  [arXiv:1005.3505 [hep-ph]];
  %%CITATION = PHRVA,D82,095006;%%
%%
%\cite{Chanowitz:2010bm}
%\bibitem{Chanowitz:2010bm}
  M.~S.~Chanowitz,
  %``Higgs Mass Constraints on a Fourth Family: Upper and Lower Limits on CKM
  %Mixing,''
  Phys.\ Rev.\  D {\bf }, 035018 (2010)
  [arXiv:1007.0043 [hep-ph]];
  %%CITATION = PHRVA,D,035018;%%
%\cite{Aslam:2010mc}
%\bibitem{Aslam:2010mc}
  M.~J.~Aslam,
  %``Semileptonic B to Scalar meson Decays in the Standard Model with Fourth
  %Generation,''
  arXiv:1007.4865 [hep-ph];
  %%CITATION = ARXIV:1007.4865;%%
%%
%\cite{Mohanta:2010eb}
%%\bibitem{Mohanta:2010eb}
  R.~Mohanta and A.~K.~Giri,
  %``Fourth generation effect on $\Lambda_b$ decays,''
  Phys.\ Rev.\  D {\bf 82}, 094022 (2010)
  [arXiv:1010.1152 [hep-ph]].
  %%CITATION = PHRVA,D82,094022;%%
%%
A.~K.~Alok, A.~Dighe and D.~London,
  %``Constraints on the Four-Generation Quark Mixing Matrix from a Fit to
  %Flavor-Physics Data,''
  arXiv:1011.2634 [hep-ph];
  %%CITATION = ARXIV:1011.2634;%%
%\cite{Nandi:2010zx}
%\bibitem{Nandi:2010zx}
  S.~Nandi and A.~Soni,
  %``Constraining the mixing matrix for Standard Model with four generations:
  %time dependent and semi-leptonic CP asymmetries in $B_d^0$, $B_s$ and
  %$D^0$,''
  arXiv:1011.6091 [hep-ph].
  %%CITATION = ARXIV:1011.6091;%%
%%
%\cite{Chen:2011te}
%%\bibitem{Chen:2011te}
  H.~Chen and W.~Huo,
  %``New physical effects on the decay $B_{s(d)} \to \gamma\gamma$ in the
  %sequential fourth Generation model,''
  arXiv:1101.4660 [hep-ph].
  %%CITATION = ARXIV:1101.4660;%%
%
%\cite{Flacco:2011ym}
%\bibitem{Flacco:2011ym}
  C.~J.~Flacco, D.~Whiteson and M.~Kelly,
  %``Fourth generation quark mass limits in CKM-element space,''
  arXiv:1101.4976 [hep-ph].
  %%CITATION = ARXIV:1101.4976;%%

\bibitem{cpviolation}
%\cite{Hou:2010wf}
%\bibitem{Hou:2010wf}
  W.~S.~Hou, Y.~Y.~Mao and C.~H.~Shen,
  %``Leading Effect of CP Violation with Four Generations,''
  Phys.\ Rev.\  D {\bf 82}, 036005 (2010)
  [arXiv:1003.4361 [hep-ph]]; 
  %%CITATION = PHRVA,D82,036005;%%
%\cite{Hou:2010mm}
%\bibitem{Hou:2010mm}
  W.~S.~Hou and C.~Y.~Ma,
  %``Flavor and CP Violation with Fourth Generations Revisited,''
  Phys.\ Rev.\  D {\bf 82}, 036002 (2010)
  [arXiv:1004.2186 [hep-ph]].
  %%CITATION = PHRVA,D82,036002
  
  
  
  


\bibitem{dphysics}
  K.~S.~Babu, X.~G.~He, X.~Li and S.~Pakvasa,
  %``FOURTH GENERATION SIGNATURES IN D0 - anti-D0 MIXING AND RARE D DECAYS,''
  Phys.\ Lett.\  B {\bf 205}, 540 (1988).
  %%CITATION = PHLTA,B205,540;%%
  %%
%\cite{Buras:2010nd}
%\bibitem{Buras:2010nd}
  A.~J.~Buras, B.~Duling, T.~Feldmann, T.~Heidsieck, C.~Promberger and S.~Recksiegel,
  %``The Impact of a 4th Generation on Mixing and CP Violation in the Charm
  %System,''
  JHEP {\bf 1007}, 094 (2010)
  [arXiv:1004.4565 [hep-ph]];
   %%CITATION = JHEPA,1007,094;%%%%
   
   
\bibitem{menkel}
%\cite{Antipin:2009ks}
%\bibitem{Antipin:2009ks}
  O.~Antipin, M.~Heikinheimo and K.~Tuominen,
  %``Natural fourth generation of leptons,''
  JHEP {\bf 0910} (2009) 018
  [arXiv:0905.0622 [hep-ph]].
  %%CITATION = JHEPA,0910,018;%%
%%
%\cite{Lacker:2010zz}
%\bibitem{Lacker:2010zz}
  H.~Lacker and A.~Menzel,
  %``Simultaneous Extraction of the Fermi constant and PMNS matrix elements in
  %the presence of a fourth generation,''
  JHEP {\bf 1007}, 006 (2010)
  [arXiv:1003.4532 [hep-ph]].
  %%CITATION = JHEPA,1007,006;%%

\bibitem{buras_lepton}
A.~J.~Buras, B.~Duling, T.~Feldmann, T.~Heidsieck and C.~Promberger,
  %``Lepton Flavour Violation in the Presence of a Fourth Generation of Quarks
  %and Leptons,''
  JHEP {\bf 1009}, 104 (2010)
  [arXiv:1006.5356 [hep-ph]].
  %%CITATION = JHEPA,1009,104;%%


\bibitem{kribs2}
R.~Fok and G.~D.~Kribs,
  %``Four Generations, the Electroweak Phase Transition, and Supersymmetry,''
  Phys.\ Rev.\  D {\bf 78}, 075023 (2008)
  [arXiv:0803.4207 [hep-ph]].
  %%CITATION = PHRVA,D78,075023;%%  

\bibitem{sher}
  S.~Litsey and M.~Sher,
  %``Higgs Masses in the Four Generation MSSM,''
  Phys.\ Rev.\  D {\bf 80}, 057701 (2009)
  [arXiv:0908.0502 [hep-ph]].
  %%CITATION = PHRVA,D80,057701;%%


\bibitem{ewbaryo}
 W.~S.~Hou,   
  %``CP Violation and Baryogenesis from New Heavy Quarks,''   
  Chin.\ J.\ Phys.\  {\bf 47}  (2009) 134   
  [arXiv:0803.1234 [hep-ph]];   
  %%CITATION = CJOPA,47,134;%%   
  W.~S.~Hou, Y.~Y.~Mao and C.~H.~Shen,  
  %``Leading Effect of CP Violation with Four Generations,''  
  arXiv:1003.4361 [hep-ph];  
  %%CITATION = ARXIV:1003.4361;%%  
  M.~S.~Carena, A.~Megevand, M.~Quiros and C.~E.~M.~Wagner,      
  %``Electroweak baryogenesis and new TeV fermions,''      
  Nucl.\ Phys.\  B {\bf 716} (2005) 319      
  [arXiv:hep-ph/0410352];    
  %%CITATION = NUPHA,B716,319;%%      
  Y.~Kikukawa, M.~Kohda and J.~Yasuda,   
  %``The strongly coupled fourth family and a first-order electroweak phase   
  transition (I) quark sector,''   
  Prog.\ Theor.\ Phys.\  {\bf 122}  (2009) 401   
  [arXiv:0901.1962 [hep-ph]].   
  %%CITATION = PTPKA,122,401;%%   

\bibitem{gvw}
R.~M.~Godbole, S.~K.~Vempati and A.~Wingerter,
  %``Four Generations: SUSY and SUSY Breaking,''
  JHEP {\bf 1003}, 023 (2010)
  [arXiv:0911.1882 [hep-ph]].
  %%CITATION = JHEPA,1003,023;%%

\bibitem{nandimurdock}
  Z.~Murdock, S.~Nandi and Z.~Tavartkiladze,
  %``Perturbativity and a Fourth Generation in the MSSM,''
  Phys.\ Lett.\  B {\bf 668}, 303 (2008)
  [arXiv:0806.2064 [hep-ph]].
  %%CITATION = PHLTA,B668,303;%%

%%%%%%%%%%%%%%%experimental limits %%%%%%%%%%%%
%\cite{Abachi:1995ms}
\bibitem{Abachi:1995ms}
S.~Abachi {\it et al.} [D0 Collaboration],
%``Top quark search with the D\O\ 1992 - 1993 data sample,''
Phys.\ Rev.\ D {\bf 52}, 4877 (1995).
%%CITATION = PHRVA,D52,4877;%%

%\cite{:2008nf}
\bibitem{:2008nf}
T.~Aaltonen {\it et al.} [CDF Collaboration],
%``Search for Heavy Top-like Quarks Using Lepton Plus Jets Events in 1.96-TeV
%$p \bar{p}$ Collisions,''
Phys.\ Rev.\ Lett.\ {\bf 100}, 161803 (2008)
[arXiv:0801.3877 [hep-ex]].
%%CITATION = PRLTA,100,161803;%%


%\cite{Achard:2001qw}
\bibitem{Achard:2001qw}
P.~Achard {\it et al.} [L3 Collaboration],
%``Search for heavy neutral and charged leptons in $e^{+} e^{-}$ annihilation
%at LEP,''
Phys.\ Lett.\ B {\bf 517}, 75 (2001)
[arXiv:hep-ex/0107015].
%%CITATION = PHLTA,B517,75;%%



\bibitem{sumit}
Garg et. al, In preparation. 



\bibitem{gabbiani}
 F.~Gabbiani, E.~Gabrielli, A.~Masiero and L.~Silvestrini,
  %``A complete analysis of FCNC and CP constraints in general SUSY extensions
  %of the standard model,''
  Nucl.\ Phys.\  B {\bf 477}, 321 (1996)
  [arXiv:hep-ph/9604387].
  %%CITATION = NUPHA,B477,321;%%


\bibitem{whepX1}
Manimala Mitra et.al : "Models for fourth generation Dirac \& Majorana neutrinos and their phenomenology", Work
presented at Whep XI, Jan. 2010 PRL, Ahmedabad, India and references there in.



\bibitem{hisanonomura}
 J.~Hisano and D.~Nomura,
  %``Solar and atmospheric neutrino oscillations and lepton flavor violation  in
  %supersymmetric models with the right-handed neutrinos,''
  Phys.\ Rev.\  D {\bf 59}, 116005 (1999)
  [arXiv:hep-ph/9810479].
  %%CITATION = PHRVA,D59,116005;%%

\bibitem{paride1}
 P.~Paradisi,
  %``Constraints on SUSY lepton flavour violation by rare processes,''
  JHEP {\bf 0510}, 006 (2005)
  [arXiv:hep-ph/0505046].
  %%CITATION = JHEPA,0510,006;%% 

%\cite{Hisano:2008hn}
\bibitem{hisanoparide}
  J.~Hisano, M.~Nagai and P.~Paradisi,
  %``Flavor effects on the electric dipole moments in supersymmetric theories: A
  %beyond leading order analysis,''
  Phys.\ Rev.\  D {\bf 80}, 095014 (2009)
  [arXiv:0812.4283 [hep-ph]].
  %%CITATION = PHRVA,D80,095014;%%


\bibitem{masina}
 I.~Masina and C.~A.~Savoy,
  %``Sleptonarium (constraints on the CP and flavour pattern of scalar lepton
  %masses),''
  Nucl.\ Phys.\  B {\bf 661}, 365 (2003)
  [arXiv:hep-ph/0211283].
  %%CITATION = NUPHA,B661,365;%%


\bibitem{seiberg}
 P.~Meade, N.~Seiberg and D.~Shih,
  %``General Gauge Mediation,''
  Prog.\ Theor.\ Phys.\ Suppl.\  {\bf 177}, 143 (2009)
  [arXiv:0801.3278 [hep-ph]].
  %%CITATION = PTPSA,177,143;%%



\bibitem{mega}
{\bf MEGA} Collaboration, M.~L. Brooks {\em et~al.}, {\it {New Limit for the
  Family-Number Non-conserving Decay $\mu^+\to e^+\gamma$}},  {\em Phys. Rev.
  Lett.} {\bf 83} (1999) 1521--1524,
  \href{http://xxx.lanl.gov/abs/hep-ex/9905013}{{hep-ex/9905013}} .

\bibitem{bellebabar}
{\bf BABAR} Collaboration, B.~Aubert {\em et~al.}, {\it {Searches for Lepton
  Flavor Violation in the Decays $\tau \to e \gamma$ and $\tau \to \mu
  \gamma$}},  {\em Phys. Rev. Lett.} {\bf 104} (2010) 021802,
  \href{http://xxx.lanl.gov/abs/0908.2381}{{0908.2381}} .


\bibitem{pdg}
 K.~Nakamura {\it et al.}  [Particle Data Group],
  %``Review of particle physics,''
  J.\ Phys.\ G {\bf 37}, 075021 (2010).
  %%CITATION = JPHGB,G37,075021;%%



\bibitem{babu-kolda}
%\cite{Babu:2002et}
%\bibitem{Babu:2002et}
  K.~S.~Babu and C.~Kolda,
  %``Higgs-mediated tau --> 3mu in the supersymmetric seesaw model,''
  Phys.\ Rev.\ Lett.\  {\bf 89}, 241802 (2002)
  [arXiv:hep-ph/0206310].
  %%CITATION = PRLTA,89,241802;%%

\bibitem{anna-andrea}
%\cite{Brignole:2003iv}
%\bibitem{Brignole:2003iv}
  A.~Brignole and A.~Rossi,
  %``Lepton flavour violating decays of supersymmetric Higgs bosons,''
  Phys.\ Lett.\  B {\bf 566}, 217 (2003)
  [arXiv:hep-ph/0304081]; 
  %%CITATION = PHLTA,B566,217;%%
%\cite{Brignole:2004ah}
%\bibitem{Brignole:2004ah}
  A.~Brignole and A.~Rossi,
  %``Anatomy and phenomenology of mu tau lepton flavour violation in the
  %MSSM,''
  Nucl.\ Phys.\  B {\bf 701}, 3 (2004)
  [arXiv:hep-ph/0404211].
  %%CITATION = NUPHA,B701,3;%%

\bibitem{paride-higgs}
%\cite{Paradisi:2005tk}
%\bibitem{Paradisi:2005tk}
  P.~Paradisi,
  %``Higgs-mediated tau --> mu and tau --> e transitions in II Higgs doublet
  %model and supersymmetry,''
  JHEP {\bf 0602}, 050 (2006)
  [arXiv:hep-ph/0508054].
  %%CITATION = JHEPA,0602,050;%%


\end{thebibliography}
\end{document}